\documentclass[12pt]{iopart}

\usepackage{color,graphicx,epsfig,rotating,rotate}
\usepackage{latexsym}
\usepackage{amssymb}
\usepackage{mathabx}
\definecolor{mygreen}{rgb}{0,0.5,0}

\newcommand{\colb}[1]{\textcolor{blue}{#1}}

\graphicspath{{fig/}}

\def\ux {\mathbf{x}}
\def\uxd {\mathbf{x}'}

\def\uu {\mathbf{u}}
\def\dho{\partial}
\def\beq {\begin{equation}}
\def\bi {\begin{itemize}}
\def\eeq {\end{equation}}
\def\ei {\end{itemize}}
\def\beqa {\begin{eqnarray}}
\def\eeqa {\end{eqnarray}}
\def\rel {R_{\lambda}}
\def\la{\langle}
\def\ra{\rangle}
\def\normr{r/\eta}

\def\epsr {\epsilon_r}

\def\epsrsq {\la\epsilon^2_r\ra}

\def\omgr {\Omega_r}

\def\omgrsq {\la\Omega^2_r\ra}

\def\prlr {\Delta p_r}
\def\pl{\Delta p}
\def\prlsq {\langle (\prlr)^2 \rangle}
\def\ou{\mathbf{\omega}}
\def\nabv{\mathbf{\nabla}}

\def\stdprl {\sigma_{\prlr}}
\def\mcalp {\mathcal{P}}

\begin{document}

\title[]{\colb{Scaling of locally averaged energy dissipation and enstrophy density in isotropic turbulence}}

\author{Kartik P Iyer$^1$, J\"{o}rg Schumacher$^{1,2}$, Katepalli R Sreenivasan$^3$, P K Yeung$^4$}

\address{$^1$Department of Mechanical and Aerospace Engineering, New York University, New York, NY, 11201, USA}
\address{$^2$Institut f\"ur Thermo-und Fluiddynamik, Technische Universit\"at Ilmenau, Postfach 100565, D-98684 Ilmenau, Germany}
\address{$^3$Institute for Advanced Study, Princeton, NJ 08540}
\address{$^4$School of Aerospace Engineering and Mechanical Engineering, Georgia Institute of Technology, Atlanta, GA 30332, USA}
\ead{kartik.iyer@nyu.edu}
\vspace{10pt}

\begin{abstract}
Using direct numerical simulations of isotropic turbulence in periodic cubes of several sizes, the largest being $8192^3$ yielding a microscale Reynolds number of $1300$, we study the properties of pressure Laplacian to understand differences in the inertial range scaling of enstrophy density and energy dissipation. Even though the pressure Laplacian is the difference between two highly intermittent quantities, it is non-intermittent and essentially follows Kolmogorov scaling, at least for low-order moments. Using this property, we show that the scaling exponents of local averages of dissipation and enstrophy remain unequal at all finite Reynolds numbers, though there appears to be a \textit{very} weak tendency for this inequality to decrease with increasing Reynolds number.  

\end{abstract}

%
\submitto{\NJP}
\vspace{1pc}
As in other highly correlated systems \cite{Harte99,christ02}, ``local" averaging over scales smaller than the system size is often employed \cite{oboukhov62,K62} in turbulence to study its statistical structure. Local averages of highly intermittent quantities are dependent on the averaging scale itself \cite{krsh,joubaud08,bramwell09} and paradigms such as the central limit theorem do not apply. Properties of local averages of energy dissipation \cite{SA97} (characterizing straining motions) and enstrophy \cite{JS2010} (characterizing local rotation) are the subject of much debate. The consensus of experimental and numerical work is that the local averages of these two quantities are different \cite{siggia1981,kerr85,MSF90,brachet91,CSN97,krsht,PK18} while theories (with some numerical support), rooted in the paradigm of small-scale universality \cite{nelkin78,He98prl,nel99,DYS2008,YDS2012,pkpnas}, conclude oppositely.

Here, we reconcile this difference by establishing two specific results. First, we show that the pressure Laplacian, which engenders the topological asymmetry between dissipation and enstrophy \cite{siggia81,she90,brachet91,VM91}, assumes a nearly self-similar (i.e., non-intermittent) form in the inertial range (IR); even though the pressure Laplacian is the difference between two highly intermittent quantities, it is non-intermittent and essentially follows Kolmogorov scaling. Second, while the pressure \cite{hm93,pumir94,vy03} does constrain the scaling of local averages of dissipation and enstrophy, the self-similar property of the pressure Laplacian implies that the exponents remain unequal at all finite Reynolds numbers, though this constraint appears to weaken very slowly with increasing Reynolds number. 

{\it{Direct Numerical Simulations}:}
We use a direct numerical simulation (DNS) database of isotropic turbulence obtained by solving the incompressible, three-dimensional Navier-Stokes equations or the components of the turbulent velocity field $u_i(\ux,t)$ with $i=x, y, z$ in a periodic cube with edge length $L_0 = 2\pi$, spanning a wide range of Reynolds numbers \cite{DY10}. Taylor microscale Reynolds numbers up to $1300$ were used. The largest DNS was conducted on a grid size of $8192^3$ \cite{pkpnas}. A statistically steady state was obtained by forcing the low Fourier modes \cite{DY10}. Averages over ten large-eddy turnover times were used for the analysis; $\la \cdot \ra$ denotes space/time averages. 

{\it{Definitions}.}
It is well known in homogeneous turbulence that $\epsilon = \Omega + 2\nu (\dho u_i/\dho x_j \; \dho u_j/\dho x_i)$, where $ \epsilon \equiv \frac{\nu}{2} (\dho u_i/\dho x_j+\dho u_j/\dho x_i)^2$
is the turbulent energy dissipation rate per unit mass, $\nu$ is the kinematic viscosity of the fluid and the summation convention is implied; the enstrophy density is given by $\Omega \equiv \nu |\ou|^2$, where $\omega = \nabv \times \uu$ is the vorticity.
\begin{figure}
\centering
\includegraphics[width=0.95\textwidth]{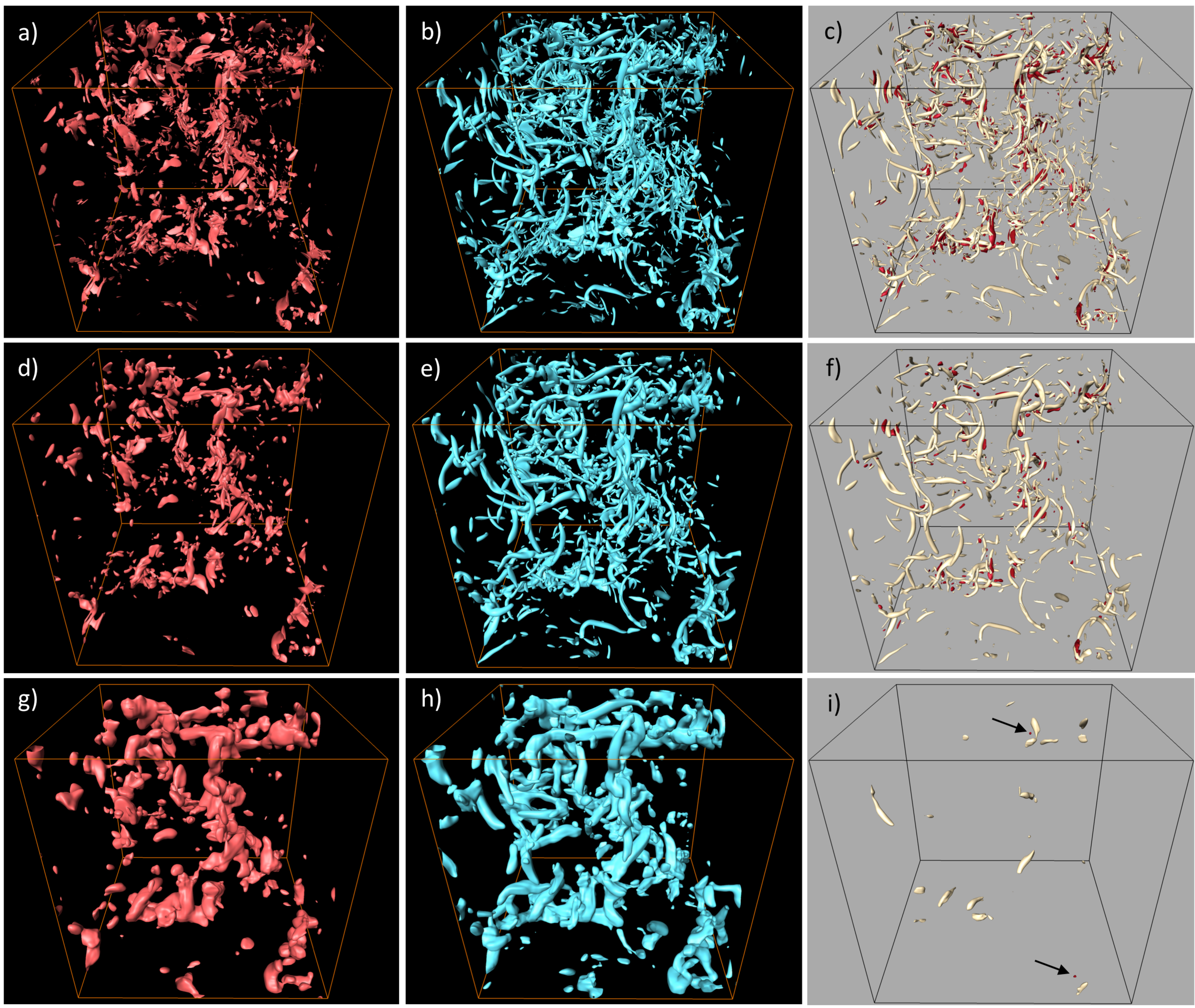}
\protect\caption{Isosurface plots of the kinetic energy dissipation rate (a,d,g), the local enstrophy (b,e,h), and the pressure Laplacian (c,f,i)). Data are obtained from the DNS at the highest Reynolds number in a cubic sub-volume with a side length $L_0/16$. Panels (a,b,c) show the data without spatial averaging at an isosurface level of $1.5$ in characteristic units of the DNS. Panels (d,e,f) display the fields for $r=10 \eta$ at a level of $1.0$; $\eta=(\nu^{3}/\langle\epsilon\rangle)^{1/4}$ is the Kolmogorov length of the flow. Panels (g,h,i) display the field at $r=33 \eta$ at a level of $0.5$. Positive isosurfaces for the pressure Laplacian are shown in yellow and negative ones in red, in panels (c,f,i). In panel (i) negative contours are indicated by black arrows.
}
\label{fields.fig}
\end{figure}

Define the local average of dissipation and enstrophy at scale $r$ as,
\beq
\label{epsr.eq}
\epsr(\ux,t) = \frac{1}{V_r} \int_{V_r} \epsilon(\ux+\uxd,t) d\uxd \; , \quad 
\omgr(\ux,t) = \frac{1}{V_r} \int_{V_r} \Omega(\ux+\uxd,t) d\uxd \;,
\eeq
where $V_r = r^3$ is a volume centered around $\ux$. Taking the divergence of the Navier-Stokes equations at constant mass density $\rho_0$, we obtain the Poisson equation for the pressure field $p$, which can then be related via its Laplacian to $\epsilon$ and $\Omega$ as $\Omega = \epsilon + 2\nu\pl/\rho_0$. Averaging this relation over volume of scale $r \ge 0$, we get 
\beq
\label{nsc.eq}
\omgr(\ux,t) = \epsr(\ux,t) + (2\nu/\rho_0) \prlr(\ux,t) \;,
\eeq 
where $\prlr$ is the locally averaged field of the pressure Laplacian $\pl$ over scale $r$ and is given by the surface integral, in accordance with Gauss's theorem, as
\beq
\label{prl.eq}
\prlr(\ux,t) \equiv \frac{2\nu\rho_0}{V_r} \int_{s_r} \frac{\dho} {\dho x_j} u_i(\ux+\uxd,t) u_j(\ux+\uxd,t) ds_i \;.
\eeq 
Here, $s_r$ denotes the surface around volume $V_r$. For brevity, in what follows, we drop the dependence on $(\ux,t)$: for instance, $\epsr \equiv \epsr(\ux,t)$. We note that in homogeneous turbulence, for any $r \ge 0$, $ \la \prlr \ra = 0$ or equivalently, $\la \epsr \ra = \la \omgr \ra$. But higher moments of $\epsr$ and $\omgr$ can differ.

{\it{Results}.}
To build up some intuition, we show in Fig.~\ref{fields.fig} all three fields in a small sub-volume of the computational box, coarse grained with respect to scale $r$. The top row shows quantities without any averaging while the bottom two rows are locally averaged quantities for two averaging scales $r$ and two different thresholds. The top row displays the typical structure: while the dissipation layers appear more sheet-like, maxima of the local enstrophy are tube-like \cite{JS2010,pkpnas}. This is a fingerprint of the local vortex-stretching, a central building block of three-dimensional turbulence. High-amplitude shear layers (leading to dissipation) are the result of self-induced strain and result in ongoing stretching \cite{PH08,PH08pof}. With increasing $r$, the differences between $\epsilon_r$ and $\Omega_r$ become less consequential, as the panels (f) and (i) show. Panels (c), (f) and (i) show that the local pressure Laplacian isosurfaces of positive amplitude appear to have a spatial correlation with local enstrophy. Pressure minima or pressure Laplacian maxima are found in the low dissipation vortex cores beyond which $\prlr$ and $\epsr$ tend to be positively correlated, since positive local pressure Laplacian isosurfaces have a greater correspondence with local dissipation at larger scales (compare panels (f,i) with (d,g), respectively), in relation to that at the smallest scales (compare panel (c) with (a)). The pressure Laplacian isosurfaces of negative amplitude (but same magnitude) are much more sparsely distributed indicating a positively skewed field.
\begin{figure}
\centering
\includegraphics[width=0.7\textwidth]{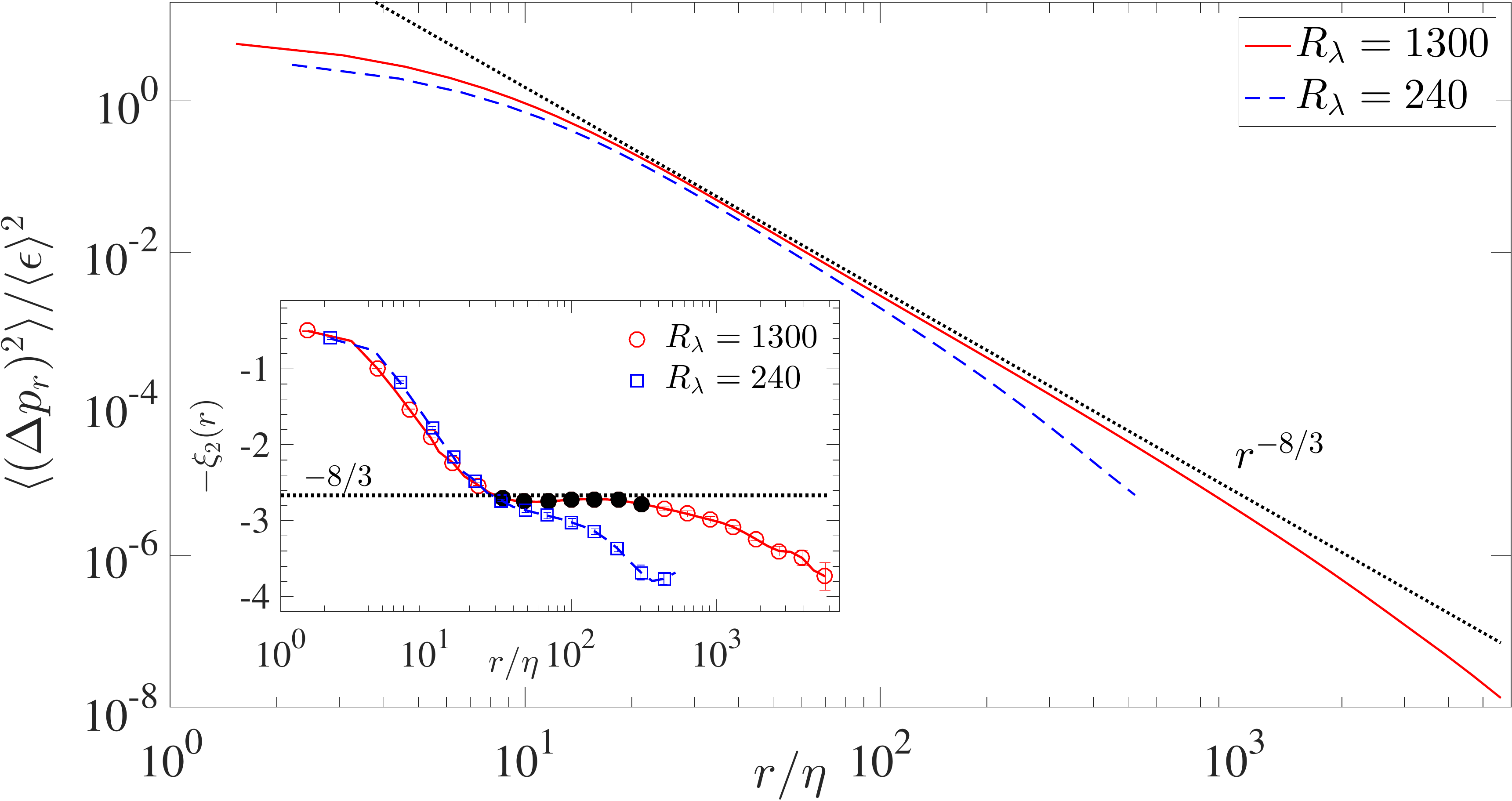}
\vspace{-0.5cm}
\protect\caption{
Variance of the locally averaged pressure Laplacian as a function of the averaging scale $r$. Dashed line corresponds to the Kolmogorov mean-field \cite{K41b} scaling of $r^{-8/3}$. Inset shows the exponent $\xi_2(r) = d[\log\prlsq]/d[\log r]$ {\it{vs}}.~$\normr$. The dotted line corresponds to the self-similar exponent, $\xi_2 = 8/3$. The IR, $\normr \in (25,350)$ for $\rel=1300$, is marked by $(\bullet)$. 
}
\label{prlvar.fig}
\end{figure}

By examining the variance, skewness and flatness of the pressure Laplacian $\prlr$, we now show that it has attributes of self-similar Kolmogorov-like scaling at high Reynolds numbers. First, Fig.~\ref{prlvar.fig} shows that the variance of $\prlr$ in IR ($\eta \ll r \ll L$) follows the power-law, $\la (\prlr)^2 \ra \sim r^{-\xi_2}$ where $\xi_2$ is the scaling exponents. Kolmogorov's arguments \cite{K41b}, which do not account for intermittency, imply that $\xi_2 = 8/3$; the measured second-order exponent is quite close (see the inset which presents local slopes), with the higher Reynolds number data showing better conformity. The trend towards intermittency-free scaling of $\prlsq$ reported here is consistent with that of the $\pl$ spectrum, reported in Refs.~\cite{gotoh01,ishihara03}.

\begin{figure}
\centering
\includegraphics[width=0.7\textwidth]{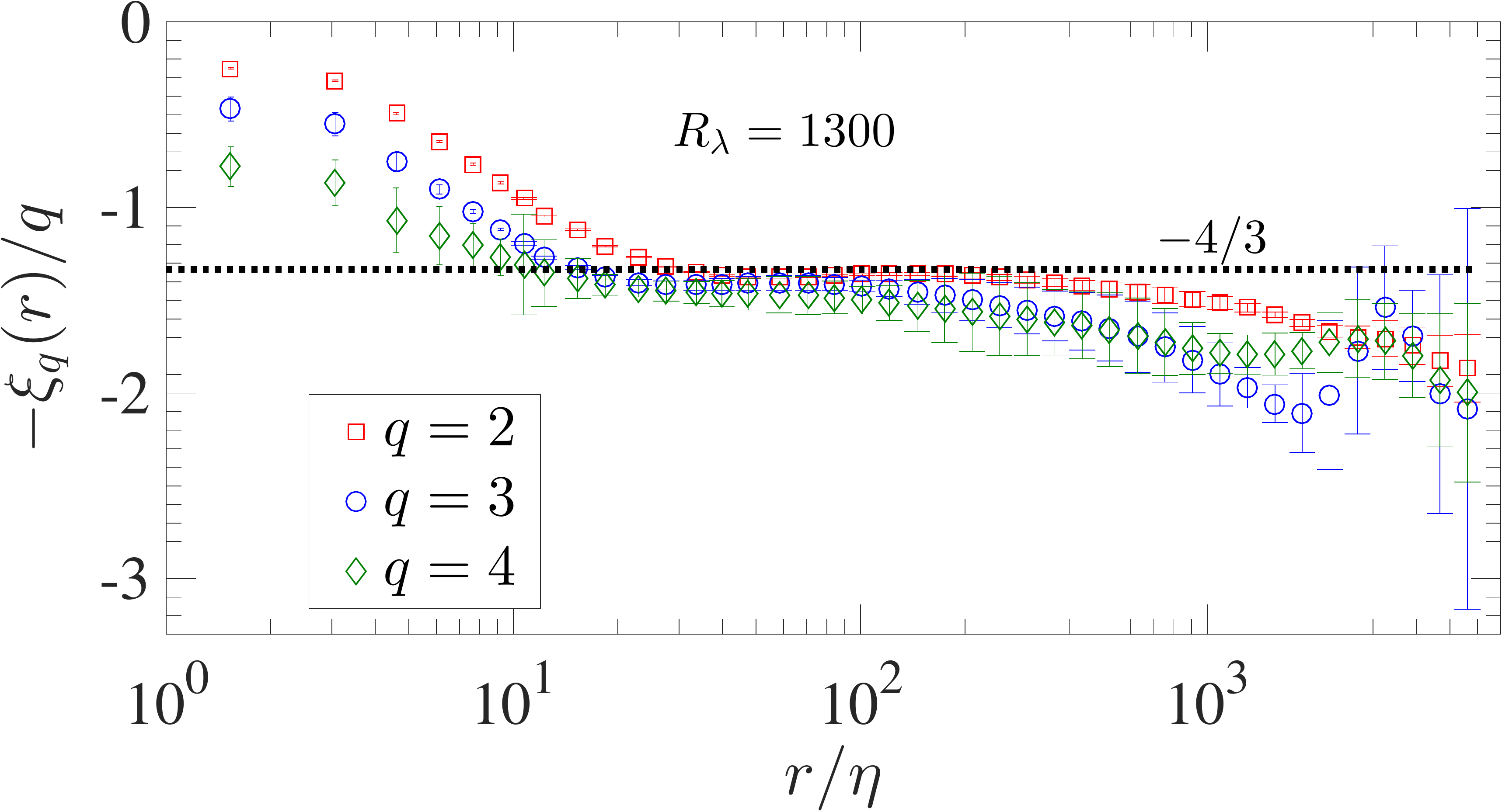}
\vspace{-0.5cm}
\protect\caption{
Quantities $\sigma_q \coloneq -\xi_q/q$ for $q = 2,3$ and $4$, plotted against the averaging scale $r$ for $R_\lambda = 1300$, where $\xi_q$ is the scaling exponent of $\prlr$ at order $q$, 
$\la (\Delta p_r)^q \ra \sim r^{-\xi_q}$. The curves are close to the self-similar Kolmogorov value of $-4/3$ (dotted line) in IR. The error bars correspond to $95\%$ confidence intervals obtained from temporal variations of the local slopes using a student's-$t$ distribution.
}
\label{ss.fig}
\end{figure}

Now, the condition for self-similarity is that the exponent, $\sigma_q \coloneq \xi_q/q$, where $\xi_q$ is the scaling exponent of $\prlr$ at order $q$, should be a constant over a range of scales; for Kolmogorov scaling, this ratio must be $4/3$. In Fig.~\ref{ss.fig}, we plot the local slopes for the second, third and fourth moments in this self-similar format for the highest Reynolds number considered. Although the behavior of the fourth order is less convincing than that of the second, there exists the tendency to the constant value of $4/3$ (within the error bars of $\sigma_q$), leading to the conclusion that $\prlr$ is likely to be a self-similar quantity following the Kolmogorov scaling at least at low orders; note, however, that the $4$th moment of $\prlr$ is equivalent to the $12$th moment for velocity differences \cite{K62}; this renders fourth-order statistics more sensitive to finite sampling effects, compared to lower orders.
\begin{figure}
\centering
\vspace{0.2cm}
\includegraphics[width=0.7\textwidth]{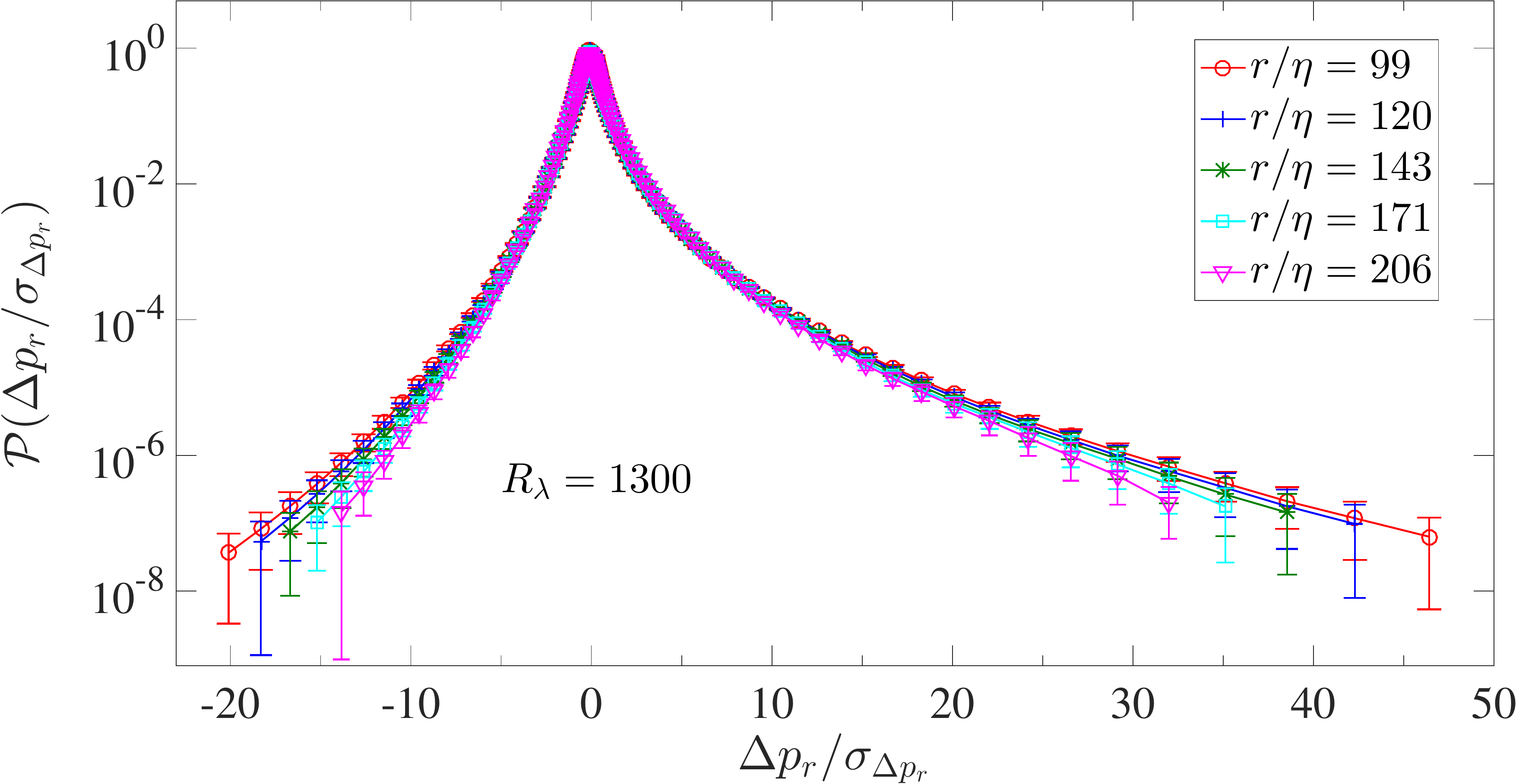}
\vspace{-0.4cm}
\caption{
Probability density function (PDF) of the scale averaged pressure Laplacian $\prlr$, normalized by its standard deviation, $\stdprl$, at $\rel = 1300$, for different IR separations. Error bars correspond to the
standard deviation of $\mcalp(\prlr/\stdprl)$ over an ensemble of $16$ temporal snapshots over 
$10$ eddy turnover times. The PDFs collapse across the IR within error bars, demonstrating the self-similar nature of $\prlr$.
}
\label{pdf_fixRe.fig}
\end{figure}

Analogously, the probability density functions (PDFs) of $\prlr$, normalized by the respective 
standard deviation, $\stdprl={\prlsq}^{1/2}$, collapse for different averaging scales $r$ in the inertial range,
as seen in Fig.~\ref{pdf_fixRe.fig}, confirming that $\prlr$ is indeed self-similar. The PDF tails collapse within error bars, but less perfectly than the bulk, possibly due to finite sampling and finite Reynolds number effects. Furthermore, the PDFs are distinctly non-Gaussian and positively skewed. The skewness increases with Reynolds number, suggesting that high enstrophy and low dissipation events $(\prlr > 0)$, are increasingly more probable than the converse $(\prlr < 0)$, over inertial length scales.
\begin{figure}
\centering
\includegraphics[width=0.7\textwidth]{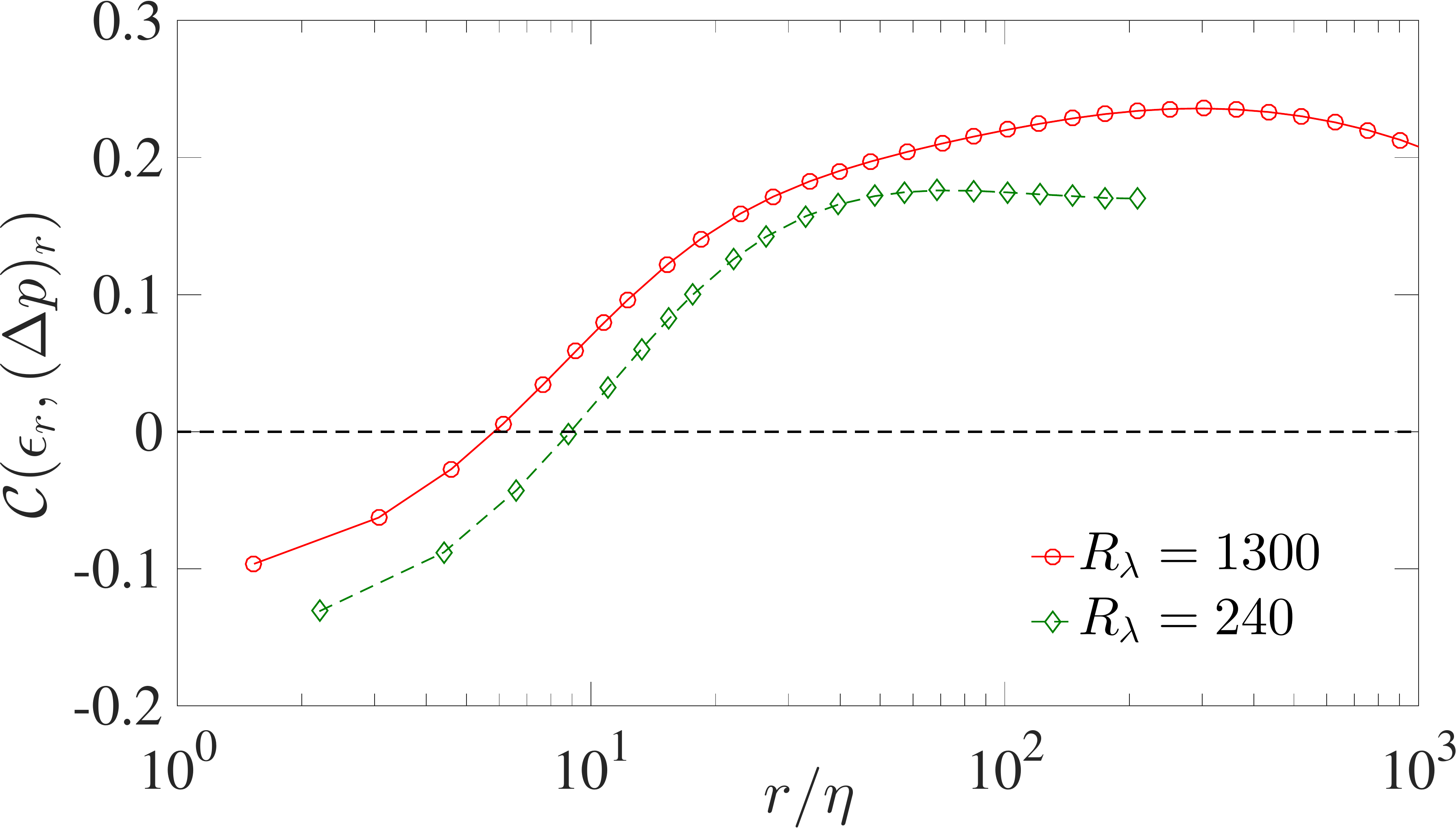}
\vspace{-0.5cm}
\caption{Correlation coefficient between scale averaged dissipation $\epsr$ and scale averaged pressure Laplacian $\prlr$, as a function of scale $r$. Dashed line at zero given for reference. For $r/\eta \gg 1$, $\la \epsr \prlr \ra > 0$ and increases with $\rel$.
}
\label{corr.fig}
\end{figure}

If the skewness of $\prlr$ is positive, as can be seen visually in Fig.~\ref{fields.fig} and more quantitatively in Fig.~\ref{pdf_fixRe.fig}, it follows that spatial averages of pressure Laplacian and dissipation will be positively correlated. In fact, all $\la \epsr^m (\prlr)^n \ra > 0$ for $m \ge 1$ and $n \ge 1$ in the inertial range. Figure \ref{corr.fig} shows, for the $m = n = 1$ case, the correlation coefficient, $\mathcal{C}(\epsr,\prlr) \equiv \la \epsr \prlr \ra/\sigma_{\epsr}\sigma_{\prlr}$. As $r/\eta \to 0$, $\la \epsr \prlr \ra < 0$, consistent with the picture that turbulence is comprised of low pressure (or high pressure Laplacian), high enstrophy vortex structures, wrapped around which are high dissipation sheets \cite{zsshe90,Douady91,VM91,pkpnas}. When the averaging scale $r$ increases, the average product $\la \epsr \prlr \ra $ eventually becomes positive since $\prlr$ is non-intermittent and remains positively skewed.
Interestingly, the zeros of the two curves in Fig.~\ref{corr.fig} occur at about $6\eta$ and $8\eta$ for $\rel=1300$ and $240$, respectively; this is of the order of the characteristic scale that can be associated with the elementary Burgers vortex stretching mechanism, the Burgers radius, $r_B \approx 4\eta$ \cite{kambe2000}. We find that the zero moves to smaller values as the Reynolds number increases and saturates at about $5\eta$, which implies that the high-dissipation shear layer is wrapped as close as possible around the core of the stretched vortex filament at higher Reynolds numbers, $R_{\lambda} \gtrsim 1000$.  
\begin{figure}
\centering
\includegraphics[width=0.7\textwidth]{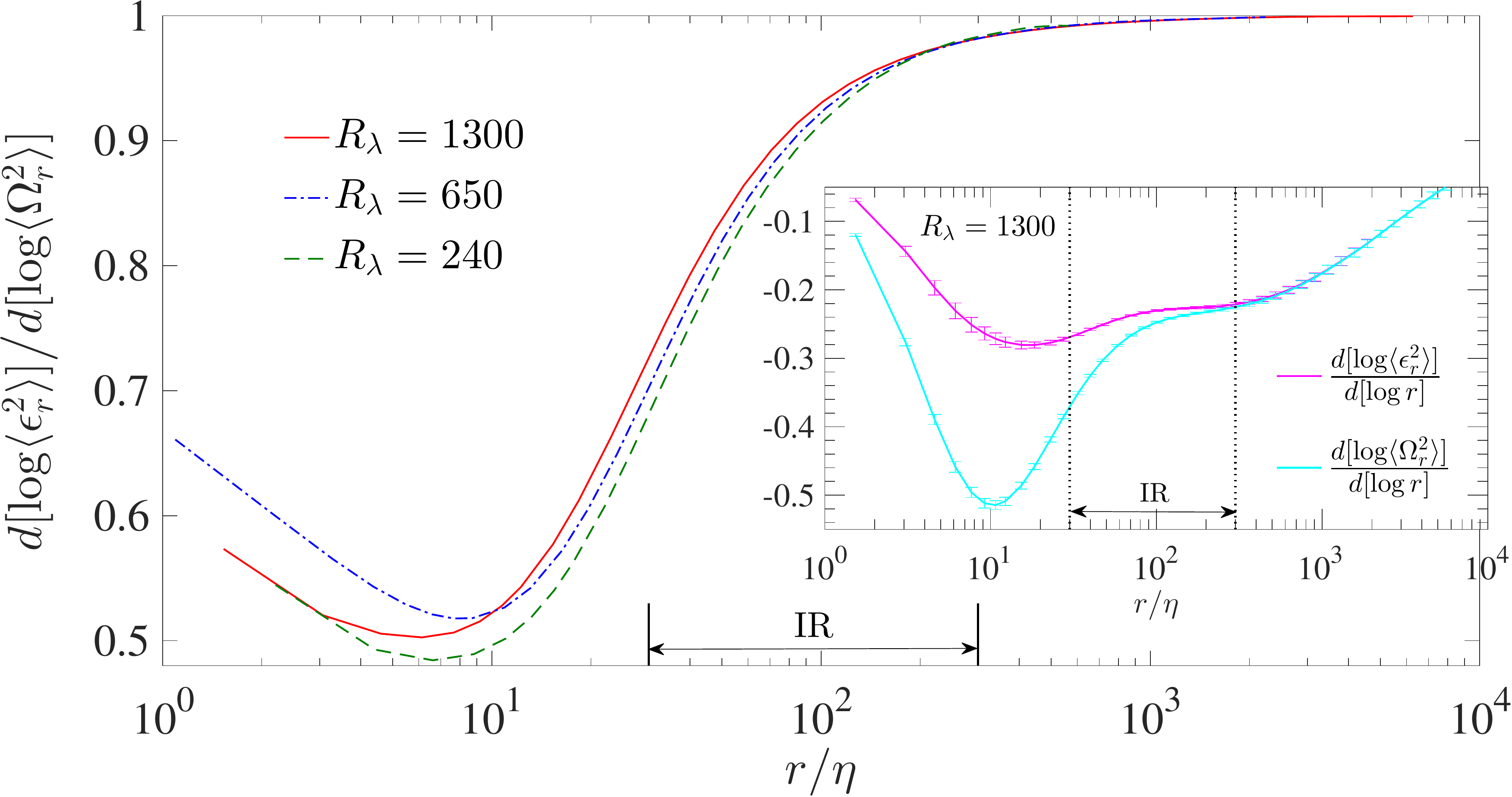}
\vspace{-0.5cm}
\caption{
Logarithmic derivative of the second order local average of dissipation $\epsrsq$ with respect to second order local average of enstrophy $\omgrsq$, as a function of the averaging scale $r$. 
The data show a very slow approach towards unity with $\rel$ in the inertial range. If the two scaling exponents have to be equal, the ordinate should be unity in IR (marked explicitly for the highest Reynolds number). The inset shows the local slopes of $\epsrsq$ and $\omgrsq$ separately at $\rel=1300$, and shows that locally averaged dissipation scales better than locally averaged enstrophy in the IR. Error bars indicate
$95\%$ confidence intervals.
}
\label{ratio.fig}
\end{figure}

To see how the positivity of $\la \epsr^m (\prlr)^n \ra$ in the IR affects the exponents of scale-averaged dissipation and enstrophy, we average the $q$th power of both sides of Eq.~\ref{nsc.eq} and get
\beq
\label{qpow.ep}
\la \omgr^q \ra - \la \epsr^q \ra = \la (\prlr)^q \ra + \sum_{m=1}^{q-1} \frac{q!}{m! (q-m)!}\la (\prlr)^m \epsr^{q-m} \ra\;.
\eeq
In IR, since $\la (\prlr)^q \ra \ge 0$ and $\la (\prlr)^m \epsr^n \ra \ge 0$ for $m+n = q$, we conclude that $\la \omgr^q\ra \ge \la \epsr^q \ra$. Now assume that $\epsr$ and $\omgr$ follow a power law scaling in IR \cite{Fri95}, $\la \epsr^q \ra \sim r^{-\mu(q)} $ and $ \la \omgr^q \ra \sim r^{-\tau(q)}$, where $\mu(q)$ and $\tau(q)$ are independent of $\rel$ for all $r$ in the inertial range. It follows that
\beq
\label{ineq.eq}
\tau(q) \ge \mu(q)\;.
\eeq
Figure \ref{ratio.fig} verifies this expectation for order $q=2$, by showing the relative logarithmic derivative of $\epsrsq$ with that of $\omgrsq$, at different $\rel$. In the inertial range, the ratio $\mu(2)/\tau(2) < 1$ for the Reynolds numbers examined, in agreement with Eq.~\ref{ineq.eq}. There appears to be a gradual trend towards unity but the trend appears to be extremely slow. We conclude that the inequality holds for all practical Reynolds numbers.

{\it{Conclusions}:}
In the preceding sections, we have shown that the pressure Laplacian $\prlr$, which is the difference of spatial averages of two highly intermittent quantities, namely the enstrophy which quantifies rotational motions and dissipation which characterizes the strain dominated motions, is a non-intermittent scale invariant quantity at high Reynolds numbers. This suggests that rotational and straining motions can be connected statistically in a relatively simple manner which is not order-dependent. Furthermore, we have established that the statistical asymmetry of $\prlr$, results in the uni-directional ordering of the scaling exponents of the moments $\la \epsr^q \ra$ and $\la \omgr^q \ra$ as $\tau(q) \ge \mu(q)$. Since the longitudinal and transverse velocity increments
can be thought of as being related to dissipation and enstrophy, respectively \cite{krsht}, it is conceivable that an analogous situation holds for velocity increments. This is an enticing prospect, since phenomenological models that are usually created for longitudinal increments, for instance \cite{menevkrs87,SL94}, can then be generalized to the velocity increment tensor in an uncomplicated manner.

\vspace{0.3cm}
\noindent{\it{Acknowledgements}:}
We thank Theodore Drivas and Victor Yakhot for useful discussions.
This work is partially supported by the National Science Foundation
(NSF), via Grant No.~ACI-$1640771$ 
at the Georgia Institute of Technology. The computations were
performed using supercomputing resources provided through
the XSEDE consortium (which is funded by NSF) at the
Texas Advanced Computing Center at the University of Texas
(Austin), and the Blue Waters Project at the National Center
for Supercomputing Applications at the University of Illinois
(Urbana-Champaign).
\\ 
\bibliographystyle{iopart-num}
\bibliography{zebib}
\end{document}